\documentclass[structabstract, longauth, regular]{aa}  

\renewcommand{\epsilon}{\varepsilon}

\usepackage{graphicx}
\usepackage{txfonts}
\usepackage[switch]{lineno}
\usepackage{setspace}
\usepackage{caption}

\usepackage{natbib}
\bibpunct{(}{)}{;}{a}{}{,} 

\newcommand{\hd}{MWC~656}
\newcommand{\agile}{AGL J2241+4454}

\begin{document}
\title{ MAGIC observations of MWC~656, the only known Be/BH system   }

%

%
\author{
J.~Aleksi\'c\inst{1} \and
S.~Ansoldi\inst{2} \and
L.~A.~Antonelli\inst{3} \and
P.~Antoranz\inst{4} \and
A.~Babic\inst{5} \and
P.~Bangale\inst{6} \and
J.~A.~Barrio\inst{7} \and
J.~Becerra Gonz\'alez\inst{8,}\inst{25} \and
W.~Bednarek\inst{9} \and
E.~Bernardini\inst{10}
% ...
}
\institute { IFAE, Campus UAB, E-08193 Bellaterra, Spain
\and Universit\`a di Udine, and INFN Trieste, I-33100 Udine, Italy
\and INAF National Institute for Astrophysics, I-00136 Rome, Italy
\and Universit\`a  di Siena, and INFN Pisa, I-53100 Siena, Italy
\and Croatian MAGIC Consortium, Rudjer Boskovic Institute, University of Rijeka and University of Split, HR-10000 Zagreb, Croatia
\and Max-Planck-Institut f\"ur Physik, D-80805 M\"unchen, Germany
\and Universidad Complutense, E-28040 Madrid, Spain
\and Inst. de Astrof\'isica de Canarias, E-38200 La Laguna, Tenerife, Spain
\and University of \L\'od\'z, PL-90236 Lodz, Poland
\and Deutsches Elektronen-Synchrotron (DESY), D-15738 Zeuthen, Germany
\and ETH Zurich, CH-8093 Zurich, Switzerland
\and Universit\"at W\"urzburg, D-97074 W\"urzburg, Germany
\and Centro de Investigaciones Energ\'eticas, Medioambientales y Tecnol\'ogicas, E-28040 Madrid, Spain
\and Institute of Space Sciences, E-08193 Barcelona, Spain
\and Universit\`a di Padova and INFN, I-35131 Padova, Italy
\and Technische Universit\"at Dortmund, D-44221 Dortmund, Germany
\and Unitat de F\'isica de les Radiacions, Departament de F\'isica, and CERES-IEEC, Universitat Aut\`onoma de Barcelona, E-08193 Bellaterra, Spain
\and Universitat de Barcelona, ICC, IEEC-UB, E-08028 Barcelona, Spain
\and Japanese MAGIC Consortium, Division of Physics and Astronomy, Kyoto University, Japan
\and Finnish MAGIC Consortium, Tuorla Observatory, University of Turku and Department of Physics, University of Oulu, Finland
\and Inst. for Nucl. Research and Nucl. Energy, BG-1784 Sofia, Bulgaria
\and Universit\`a di Pisa, and INFN Pisa, I-56126 Pisa, Italy
\and ICREA and Institute of Space Sciences, E-08193 Barcelona, Spain
\and Universit\`a dell'Insubria and INFN Milano Bicocca, Como, I-22100 Como, Italy
\and now at: NASA Goddard Space Flight Center, Greenbelt, MD 20771, USA and Department of Physics and Department of Astronomy, University of Maryland, College Park, MD 20742, USA
% ...
}
\date{Received ... , 2014; accepted ... , 2015}

\abstract
{MWC 656 has recently been established as the first observationally detected high-mass X-ray binary system containing a Be star and a black hole (BH). The system has been associated with a gamma-ray flaring event detected by the \textit{AGILE} satellite in July 2010.} 
{Our aim is to evaluate if the \hd\ gamma-ray emission extends to very high energy (VHE > 100~GeV) gamma rays.}
{We have observed \hd\ with the MAGIC telescopes for $\sim$23 hours during two observation periods: between May and June 2012 and June 2013. During the last period, observations were performed contemporaneously with X-ray (\textit{XMM-Newton}) and optical (STELLA) instruments.}
{We have not detected the \hd\ binary system at TeV energies with the MAGIC Telescopes in either of the two campaigns carried out. Upper limits (ULs) to the integral flux above 300 GeV have been set, as well as differential ULs at a level of $\sim5\%$ of the Crab Nebula flux. The results obtained from the MAGIC observations do not support persistent emission of very high energy gamma rays from this system at a level of 2.4\% the Crab flux.}
{}

\maketitle

\section{Introduction}

High-mass X-ray binaries (HMXBs) are systems composed of a massive star ($M_{\star}$ $\geq$  10 $M_{\odot}$) and a compact object, either a black hole (BH) or a neutron star (NS).  The search for GeV and TeV emission from HMXBs has been the aim of extensive studies during the past few decades. Despite the large number of observations devoted to the search, only a few of those systems have been confirmed as gamma-ray emitters. A particular group of five systems are regularly detected at TeV energies: the gamma-ray binaries (see \citealt{Dubus13} and references therein). Two other HMXBs have been the object of extensive searches: Cygnus X-3, which emits in the high-energy (HE; 100 MeV < E < 100 GeV) domain \citep{Fermi_CygX3, Tavani2009}
, and Cyg X-1, which has been reported to emit at HE \citep{2013ApJ...766...83S, Malyshev2013, Bodaghee2013} and showed a $\sim 4\sigma$ excess at VHE \citep{Albert2007}. To investigate the gamma-ray mechanisms in this type of sources, observational campaigns on other HMXBs have been carried out. The recently discovered object \hd\ \citep{Lucarelli10} is a HMXB system and has been proposed as a new gamma-ray binary candidate \citep{Williams10}.



On July 2010, \textit{AGILE} \citep{AGILE09} detected a gamma-ray point-like source dubbed \agile\ with a significant excess above 5 sigma, displaying an integral flux above 100 MeV of  $15 \times 10^{-7}$ ph cm$^{-2}$ s$^{-1}$ \citep{Lucarelli10}.  The source was first detected during the period between 25th July at 01:00 UT (MJD = 55402.042) and 26th July 2010 at 23:30 UT (MJD  = 55403.979). The source is located at $(l, b) = (100.0^{\circ}, -12.2^{\circ}) \pm 0.6^{\circ} (95\% {\rm stat.})\pm0.1^{\circ}({\rm syst.})$. At the time of writing, no further flares from this source have been reported and no spectrum has been published.

\textit{Fermi}-LAT \citep{Fermi09} could not confirm the detection by \textit{AGILE} and an analysis\footnote{\tt http://fermisky.blogspot.com.es/2010/07/ extra-note-july-30-2010.html} of simultaneous data from the same direction yielded an upper limit (UL) of $10^{-7}$ ph cm$^{-2}$ s$^{-1}$ (95$\%$ confidence level, from now on CL) above 100 MeV, assuming a photon index $\Gamma=2$.  A more extended analysis of \textit{Fermi}-LAT data, including 3.5 years of data on the \agile\ source location, also led to no evidence of HE gamma-ray emission. A 90$\%$ CL UL was set at the level of $9.4 \times 10^{-10}$ ph cm$^{-2}$ s$^{-1}$ for 3.5 years of observations \citep{MoriHD}.


The Be star \hd, also known as HD 215227, lies within the error bars of the \textit{AGILE} best-fit source position. It was proposed as the optical counterpart of the excess claimed by the \textit{AGILE} collaboration \citep{Williams10}. The system displays optical photometric modulation with a period of 60.37 $\pm$  0.04 days \citep{Williams10,Paredes-Fortuny12}. Optical spectroscopic measurements of MWC~656 confirmed its binary nature \citep{Casares_optical_2012}.
Recent optical spectroscopic measurements improved the spectral classification and reduced the uncertainties in the spectrophotometric distance, placing the system at a distance of $2.6\pm0.6$~kpc. These measurements also revealed that the compact object is a stellar-mass BH of 3.8--6.9 solar masses, making this the first known case of a Be/BH system \citep{Casares14}. 

\hd\ was also observed in radio with the European VLBI Network (EVN) and was not detected: \cite{moldon_thesis} reported 3$\sigma$ radio flux density ULs at 30--66 $\mu$Jy level.

X-ray observations were performed by \textit{XMM-Newton} when the source was at an orbital phase $\phi =~$0.08\footnote{Phase 0 has been set to the maximum of optical brightness, on HJD 2453243.3 (MJD 53242.8). With the ephemeris from \cite{Casares14}, the periastron passage occurs at phase 0.01 $\pm$ 0.10.}  \citep{munaradrover14}. The X-ray flux measured was compared with the radio ULs, resulting in a ratio compatible with the correlation derived in \cite{Corbel13} for BH LMXBs, and comparable to the faintest BH LMXBs detected. A search for hard X-ray emission has been conducted with \textit{INTEGRAL} \citep{Li2013} with no positive detection in the 18--60 keV energy band reported for a  total exposure time of 2.1~Ms.

In addition, the \textit{MAXI} mission, which continouosly monitors the X-ray sky in the 2--20 keV band, has not detected emission coming from the \agile\ position\footnote{\tt http://maxi.riken.jp/top/index.php?cid=1\&jname= J2242+447\#lsp} on the same date as of the \textit{AGILE} detection. 

In this work we present the results of the observations of \hd\ carried out with the MAGIC telescopes in 2012 and 2013. X-ray and optical observations were performed during the 2013 campaign to study the behavior of the source in a multiwavelength context. 

\section{Observations}

VHE observations of \hd\ were carried out using the MAGIC telescopes, which are located at the observatory of El Roque de Los Muchachos (28$^\circ$N, 18$^\circ$W, 2200 m above the sea level) 
on the island of La Palma, Canary Islands, Spain. The system consists of two 17 m diameter Imaging Atmospheric Cherenkov Telescopes (IACTs) each one with a pixelized camera containing photo-multipliers, covering a field of view of 3.5$^{\circ}$. The current sensitivity of the MAGIC stereoscopic system is  0.71$\%$ $\pm$ 0.02$\%$ of the Crab Nebula flux in 50 h of observation for energies above 250 GeV \citep{Upgrade2014}. The spatial resolution at these energies is $\lesssim 0.1^\circ$ and the energy resolution is $\sim$18\%. In the case of monoscopic observations (also referred as mono observations) the integral sensitivity above 280 GeV is about 1.6\% of the Crab Nebula flux in 50 hours \citep{aliu09}. The observations are performed using wobble mode, in which the telescopes point at two different symmetric regions situated $0.4^{\circ}$ away from the source position to simultaneously evaluate the background. 

\begin{figure}[t!]
\resizebox{1.0\hsize}{!}{\includegraphics{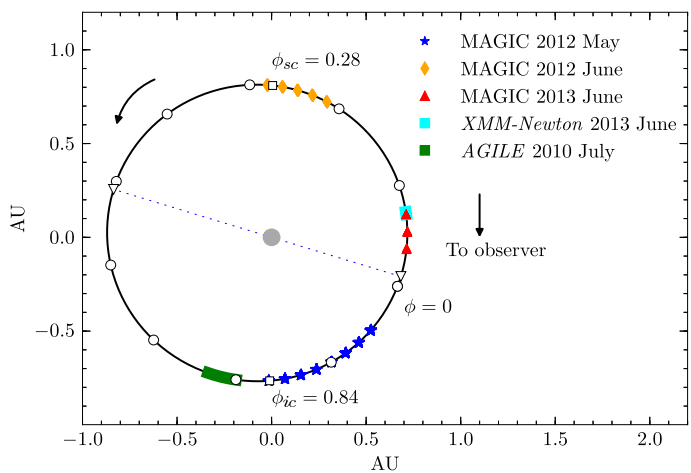}}
\caption[Sketch of the orbit of MWC~656]{Depiction of the orbit of the MWC~656 system as seen from above. The optical star MWC~656 lies at the focus of the ellipse and the BH follows the elliptical orbit. The size of the star is scaled with respect to the BH's orbit. The MAGIC, \textit{XMM-Newton} and \textit{AGILE} observations are overlaid along the orbit. Circles represent steps of 0.1 orbital phases while triangles mark the periastron and apastron phases, which are linked by a dotted line. Squares mark the position of the inferior and superior conjunctions.\label{fig:orbita_hd}}
\end{figure}

The observations of \hd\  were performed during two different epochs:  May-June 2012 and June 2013. The observations in 2012 were performed between 23rd of May and 19th of June in mono mode with MAGIC-II (the MAGIC-I telescope was not operational) for 23.4 hours.  After selecting good-quality data, a total of 21.3 hours remained. The 2013 observations were performed between June 3rd and 5th, just after the periastron passage (see Figure \ref{fig:orbita_hd}) in stereo mode. The source was observed for a total of $\sim$3.3h during this period. A summary of the observations is shown in Table~\ref{tab:hd_observations}.

\begin{table}[t!]\footnotesize

\begin{center}
   \caption{Observations of \hd\ performed by MAGIC in 2012 and 2013.  \label{tab:hd_observations}}
\begin{tabular}{c c c c c}
\hline
\hline
Date & Orbital & Zenith Angle & Time & Mode\\
(MJD) & Phase ($^{\circ}$) & Range  ($^{\circ}$)  & (hours)\\
\hline
56070 - 56078 & 0.83 - 0.95 &  23 - 50 & 9.4 & mono\\
56092 - 56097 & 0.20 - 0.28 &  22 - 51 & 14.0 & mono\\
56446 - 56448 & 0.06 - 0.08 &  28 - 45 & 3.3 & stereo\\
\hline
\end{tabular}
\end{center}

\end{table}

The observation of June 4, 2013 ($\phi$ = 0.08) lasted for $\sim 1$~h and was taken almost simultaneously with a \textit{XMM-Newton} observation (\textit{XMM-Newton started observing right after MAGIC finished its observations)}, the results of which are reported in \cite{munaradrover14}. MWC 656 was also observed with the fiber-fed STELLA  Echelle Spectrograph (SES) of the 1.2m robotic STELLA-I (ST) optical telescope \citep{stella2004} at the Observatorio del Teide in Tenerife on the nights of 2nd, 3rd, 5th and 8th of June 2013. The spectra cover the wavelength range 3870--8800 \AA ~with increasing inter-order gaps starting at 7200 \AA. The spectrograph provides an effective resolving power of $R = $55000. Two spectra were obtained on the nights of 2 and 5 June and one on the nights of 3 and 8 June. The integration time was set to 1800 s per spectrum while the automatic pipeline products were used for the extraction and calibration of the spectra.


\section{Data Analysis}

The MAGIC data analysis was performed using the standard MAGIC analysis and reconstruction software, MARS \citep{Zanin13}. The recorded shower images were calibrated, cleaned and parametrized (\citealp{1985ICRC....3..445H, aliu09}). The $\gamma$/hadron separation (background rejection) is performed via the Random Forest (RF) method \citep{magic:RF}. The event direction and the energy of the primary gamma ray were reconstructed, in the case of mono observations, also by using a RF method. The energy of each event in the case of stereoscopic observations is estimated  using look-up tables generated by Monte-Carlo simulations \citep{MAGIC_stereo_performance}. Upper limits (ULs) were derived using the method explained in \cite{Rolke2005} with a CL of 95$\%$ and a systematic uncertainty of 30\%, assuming different photon indexes ($\Gamma$ = 2.0, 2.5 and 3.0). The values obtained for the three spectral indexes are compatible at the 5\% level. The results reported in this paper are for $\Gamma=2.5$.

\section{Results}

No significant gamma-ray emission was detected from \hd\ in either observational campaign. Furthermore, no significant signal was detected in a day-to-day analysis. 

We computed 95\% confidence level (CL) integral flux ULs above 300 GeV. The integral flux UL for the whole observational campaign of \hd\ is $F(E > 300~{\rm GeV})=2.0 \times 10^{-12}$ cm$^{-2}$ s$^{-1}$ ($\sim$2.4$\%$ of the Crab Nebula flux at the same energy and corresponding to a luminosity $L_{\rm VHE} \sim {10}^{33} $erg s$^{-1}$) at 95$\%$ CL, assuming a power-law model with a photon index $\Gamma=2.5$. 

We divided the observational periods into four different phase bins, using a bin width of 0.1 in phase, along with the most recent ephemeris \citep{Casares14}. The phase was binned as follows: phases 0.8--0.9, 0.9--1.0 and 0.2--0.3 for the 2012 campaign, covering the orbit before the periastron passage and also post-periastron phases (see Figure~\ref{fig:orbita_hd}). The 2013 campaign covered the phase range 0.0--0.1,  just after the periastron passage. No signal was detected and integral ULs for bins of $\sim$0.1 in phase have also been computed. (See Table \ref{tab:phaseogram})

\begin{table}[t!]

\begin{center}
   \caption[Phase-to-phase integral ULs of \hd\ for energies above 300 GeV (95\% CL) ]{Integral flux ULs for $E>300$ GeV calculated at 95\% CL for \hd\  for each orbital phase range.  \label{tab:phaseogram}}
\begin{tabular}{c c c c c}
\hline
\hline
Mode & Phase bin & Integral UL   & Significance & t$_{eff}$\\
     &           & ($E>$300 GeV) &              &          \\
     &           & ($10^{-12}$ cm$^{-2}$ s$^{-1}$) & ($\sigma$)  & (h)    \\
\hline
stereo & 0.0--0.1 & 2.0 &  1.0 & 3.3\\
mono   & 0.2--0.3 & 8.7  &  2.1 & 4.9\\
mono   & 0.8--0.9 & 6.5  &  1.0 & 11.5\\
mono   & 0.9--1.0 & 2.5  & $-$1.1& 4.9\\
\hline
\end{tabular}
\end{center}

\end{table}

We have computed differential flux ULs from the energy threshold of our analysis (245 GeV) up to 6.3 TeV at 95$\%$ CL, with five bins per decade of energy (see Figure \ref{fig:sed}).

The MAGIC observations carried out on June 4, 2013 were performed almost simultaneously with an \textit{XMM-Newton} observation. The detected low X-ray flux was consistent with the source being in the quiescent state (defined in terms of the Eddington luminosity, when L < $10^{-5} L_{\rm Edd}$) during the observation \citep{munaradrover14}. The MAGIC integral flux UL for June 4 is  $F(E>300 {\rm GeV}) < 4.9\times10^{-12}$ cm$^{-2}$ s$^{-1}$. There is no specific information about the X-ray state of the binary system during the 2012 observations. Other space missions such as \textit{MAXI} have not reported emission from \hd\ during the 2012--2013 campaign, which might be indicative of a quiescent state as well.

Finally, the STELLA spectra, contemporaneous with the 2013 MAGIC campaign, shows the presence of the double peaked He II $\lambda$4686 emission line with an equivalent width comparable to that reported in \cite{Casares14}. We also detect other emission lines, mainly H$\alpha$, H$\beta$ and weak FeII lines with comparable strength to that measured by \cite{Casares_optical_2012}. Therefore, we conclude that MWC 656 is in a similar optical state as in past observations, the 2013 X-ray observations indicate a quiescent state, and hence the accretion activity should be very similar.

\section{Discussion and Conclusions}

We have searched for a VHE counterpart of the only known Be/BH binary system, \hd. The VHE observations performed by MAGIC can exclude a VHE flux based on the extrapolation of the emission from the \textit{AGILE} detection. Assuming a power-law spectrum and a photon index $\Gamma=2.5$, this emission would be $\sim 4\times10^{-11}$ TeV$^{-1}$ cm$^{-2}$ s$^{-1}$ at 300 GeV ($L_{\rm VHE} \sim 2 \times 10^{34}$ erg s$^{-1}$), which is well above the UL imposed by MAGIC. However, no flaring episodes were reported during the MAGIC observations, limiting the conclusions we can derive from the HE/VHE comparison.

In this type of binary, several mechanisms have been proposed that would result in gamma-ray emission above tens of GeV (\citealp{{RemillardMcClintock06}, {Fender2006},{Zdziarski2014}}). Unfortunately, the lack of contemporaneous data at other wavelengths during the \textit{AGILE} flare make conclusions on the type of emission model highly speculative.
It is even possible that the \textit{AGILE} detection was just a transient event of an unknown nature in the direction of the binary system but not related to it. Nevertheless, different emission levels could be expected depending on the state where the system is, i.e; quiescence or accreting state.


\begin{figure}[t!]
\centering
\resizebox{\hsize}{!}{\includegraphics{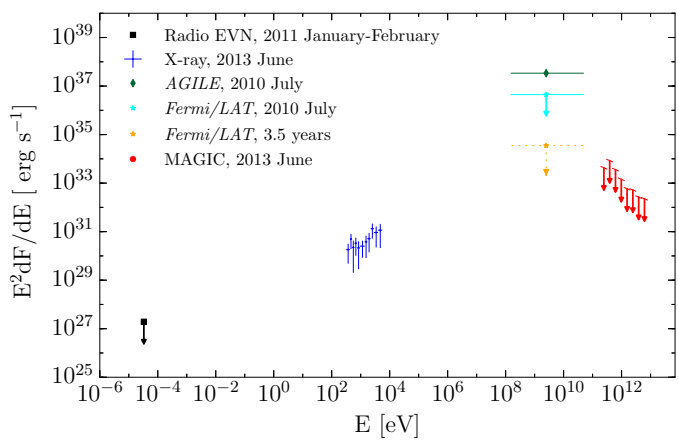}}
\caption[SED of \hd]{SED of \hd\ including MAGIC ULs from the 2013 campaign together with simultaneous \textit{XMM-Newton} data from \cite{munaradrover14}. We also include EVN radio ULs from \cite{moldon_thesis}, the \textit{AGILE} energy flux from \cite{Lucarelli10} and the \textit{Fermi}-LAT UL simultaneous (green) to the \textit{AGILE} measurement. The 3.5-year UL set by \textit{Fermi}-LAT to any persistent emission is also plotted.\label{fig:sed}}
\end{figure}

During simultaneous X-ray and VHE observations the X-ray luminosity of the source in the 0.3 -- 5.5~keV energy range was $L_{\rm X}(0.3{\rm -}5.5~{\rm keV}) = \left(1.6^{+1.0}_{-0.9}\right)\times10^{31}$ erg s$^{-1} \equiv (3.1\pm 2.3)\times10^{-8}L_{\rm Edd}$ \citep{munaradrover14} for the estimated BH mass range 3.8--6.9 $M_{\odot}$ \citep{Casares14}. The low X-ray luminosity is characteristic of systems in quiescent states (defined in terms of the Eddington luminosity, when $L_{\rm X} < 10^{-5}L_{\rm Edd}$, \citealt{Plotkin13}). For instance, the X-ray luminosity is $\sim$5 orders of magnitude lower than the one typically observed in Cygnus~X-1, which has also been observed by MAGIC \citep{Albert2007}. 
Even if we consider an increase in the X-ray flux consistent with a flaring state in the 2012 observations, and a ratio between X-rays and VHE gamma rays of $F_{\rm X}/F_{\rm TeV} \sim 10$ (similar to that observed for Cygnus~X-1 in \citealt{Albert2007}), the expected VHE emission would be a factor $\sim1.5\times10^{-5}$ of the Crab Nebula flux ( $\sim$ $2\times 10^{-15}$ cm$^{-2}$ s$^{-1}$ ) in this case, also well below the detectable levels for the current IACTs and even for the next generation of Cherenkov telescopes: the Cherenkov Telescope Array (CTA). The integral sensitivity of CTA is foreseen to reach $\sim 3\times10^{-13}$ TeV cm$^{-2}$ s$^{-1}$ above 50 GeV for 50~h of observation ($\sim 7\times10^{-12}$ TeV cm$^{-2}$ s$^{-1}$ in 1~h observation, considering array E configuration) \citep{Acharya13}, still not enough to detect \hd\ in relatively short time in this simple approximation.

Figure~\ref{fig:sed} shows the spectral energy distribution (SED) of \hd. The absence of a detection of steady emission at high and very high energies implies that the X-ray emission cannot continue to increase with energy indefinitely and must turn over in the SED. The MAGIC differential ULs correspond only to June 2013 data, because for the 2012 observations X-ray information is not available. We also plotted the \textit{AGILE} measurement in the SED along with the \textit{Fermi}-LAT upper limit, obtained with observations performed on the same dates of the \textit{AGILE} detection. Although the \textit{Fermi}-LAT UL contradicts the \textit{AGILE} detection, it is worth noting that the observation modes of these telescopes are different and that the integration time might not be exactly the same. Therefore, the observations are not strictly simultaneous and \textit{Fermi}-LAT might have missed short ( < 1 hour) gamma-ray flares from \hd.




\begin{acknowledgements}
We would like to thank the Instituto de Astrof\'{\i}sica de Canarias for the excellent working conditions at the Observatorio del Roque de los Muchachos in La Palma. The support of the German BMBF and MPG, the Italian INFN,  the Swiss National Fund SNF, and the ERDF funds under the Spanish MINECO is gratefully acknowledged. This work was also supported by the CPAN CSD2007-00042 and MultiDark CSD2009-00064 projects of the Spanish Consolider-Ingenio 2010 programme, by grant 127740 of the Academy of Finland, by the Croatian Science Foundation (HrZZ) Project 09/176, by the DFG Collaborative Research Centers SFB823/C4 and SFB876/C3, and by the Polish MNiSzW grant 745/N-HESS-MAGIC/2010/0. J.C. acknowledges support by the Spanish Ministerio de Econom\'ia y Competividad (MINECO) under grant AYA2010-18080. The authors thank the anonymous referee for a thorough review and a very constructive list of remarks that helped improving the quality and clarity of this manuscript.

\end{acknowledgements}


\bibliography{hd215227_v2_ref_2}
\bibliographystyle{aa}

\end{document}